\documentstyle[preprint,aps,epsf,floats,tighten]{revtex} 

\begin{document}

\preprint{\tighten \vbox{\hbox{   }
\hbox{} \hbox{} \hbox{} \hbox{} \hbox{} } }

\title{Semi-inclusive nonleptonic decays $B\to D_s^{(*)} X$}

\author{Changhao Jin}

\address{
School of Physics,
University of Melbourne, Victoria 3010, Australia}

\maketitle

{\tighten
\begin{abstract}%
We calculate the total and differential decay rates for the semi-inclusive 
nonleptonic decays $B\to D_s^{(*)} X_q$ ($q=c$ or $u$) under the factorization
hypothesis. The initial bound state effect is treated using the light-cone
expansion and the heavy quark effective theory. We investigate the
contribution of the penguin amplitude for the decay mode
$B\to D_s^{(*)} X_c$ and find that it is small but non-negligible. The
resulting decay rates for $B\to D_s^{(*)} X_c$ agree with the measurements.
The use of the decay mode $B\to D_s^{(*)} X_u$ to determine
$|V_{ub}|$ is discussed.
\end{abstract}
}

\newpage

\section{Introduction}
The semi-inclusive nonleptonic decays $B\to D_s^{(*)} X_q$ ($q=c$ or $u$), 
where $X_q$ is a hadronic system containing a $q$-quark, can be used to 
study mechanisms of the production of the $D_s^{(*)}$ meson in $B$ meson 
decays, as well as the dynamics of the strong interactions \cite{palmer}. 
As pointed out in Ref.~\cite{aleksan}, the Cabibbo-Kobayashi-Maskawa (CKM) 
matrix element $|V_{ub}|$ can be determined by the 
measurement of the momentum spectrum of the $D_s^{(*)}$ meson in 
$B\to D_s^{(*)} X_u$. An advantage of the method\footnote{Other methods of
extracting $|V_{ub}|$ from nonleptonic $B$ decays have also been 
proposed \cite{nonlep}.} 
is that the majority of 
the spectrum in the rare decays $B\to D_s^{(*)} X_u$ lie above the kinematic 
limit for $B\to D_s^{(*)} X_c$ which constitutes an overwhelming background.  
Since this method involves
nonleptonic $B$ decays, in which only hadrons appear in the final state, 
the hadronic complications may be more sever than the
determination of $|V_{ub}|$ from semileptonic decays 
$B\to \ell\bar\nu_\ell X_u$. An understanding of the strong interactions,
which are responsible for the confinement of quarks and gluons in hadrons,
in the weak decays is essential to a determination of $|V_{ub}|$.
  
Measurements of the branching fraction for $B\to D_s X$ have been 
reported by ARGUS and CLEO. The average of their results gives 
${\cal B}(B\to D_s^\pm X)=(10.0\pm 2.5)\%$ \cite{PDG}. Recently,
new precise measurements of $D_s^+$ and $D_s^{*+}$ meson production
from $B$ meson decays have been presented by BABAR \cite{babar}. The 
results for the branching fractions are
${\cal B}(B\to D_s^+ X)=(10.93\pm 0.19\pm 0.58\pm 2.73)\%$ and
${\cal B}(B\to D_s^{*+} X)=(7.9\pm 0.8\pm 0.7\pm 2.0)\%$.  
As for the rare decays $B\to D_s^{(*)} X_u$, none of them have been 
experimentally measured. With
high-statistics event sample from the $B$ factories, coupled with 
improvements in detector performance and analysis
techniques, measurements of 
$B\to D_s^{(*)} X_u$ might be possible in the near future.

The $b\to q\bar c s$ tree transition is expected to dominate  
$\bar B\to D_s^{(*)-} X_q$ decays. In these processes the $b$ quark decays to
a $c$ or a $u$ quark, emitting a virtual $W$ that fragments to a 
$\bar c s$ quark pair that subsequently hadronizes into a $D_s^{(*)-}$ meson. 
The effective weak Hamiltonian responsible for the decays is well 
known \cite{BBL}, which is based on the operator product 
expansion \cite{wilson} and the renormalization group \cite{rg}. 
The problem is how to calculate hadronic matrix 
elements of four-quark operators in the effective weak Hamiltonian. 
Factorization hypothesis has widely been used in calculations of nonleptonic
$B$ decays, which reduce the
hadronic matrix element of a four-quark operator to the product of two 
matrix elements of current operators \cite{fact}. 
The argument for factorization is based on the space-time evolution of the
decay products \cite{bj}. At the quark level the decay 
$\bar B\to D_s^{(*)-} X_q$ begins as a nearly collinear configuration of
$b\to q + (\bar c s)$, with $q$ and the $(\bar c s)$ pair moving 
rapidly apart in opposite directions. When the color singlet $D_s^{(*)-}$ 
meson is formed,
it is far away from the hadronic system $X_q$ recoiling against it. 
The strong interaction between them is expected to be not 
significant. Therefore, the hadronic
matrix element can be factorized into a product of hadronic matrix elements 
of color-singlet quark currents. For $\bar B\to D_s^{(*)-} X_c$ due to 
limited energy release this argument is considered to be weaker than 
$\bar B\to D_s^{(*)-} X_u$. Factorization in the nonleptonic $B$ decays with 
the emission particle being a light
meson can be justified more rigorously in the heavy quark 
limit using perturbative QCD, where the emission particle 
means the one that does not inherit the spectator quark from the $B$ 
meson \cite{kls,bbns,bps}. 
Unfortunately, this is not the case for both $\bar B\to D_s^{(*)-} X_c$ and 
$\bar B\to D_s^{(*)-} X_u$ decays, where the emission particle is $D_s^{(*)-}$
that is not light. Nevertheless, factorization has been
found to hold with current accuracy in the color-favored decay modes
$\bar{B^0}\to D_s^{(*)-}D^{(*)+}$, where no perturbative QCD justification 
has been presented \cite{luo}. Thus, we have some confidence in the
applicability of factorization to the color-favored decay modes
$\bar B\to D_s^{(*)-} X_q$.

In this paper, we analyze $\bar B\to D_s^{(*)-} X_q$ decays in some detail.
We shall calculate the total and differential $\bar B\to D_s^{(*)-} X_q$ 
decay rates assuming factorization. We include the contributions of penguin 
diagrams to $\bar B\to D_s^{(*)-} X_c$ decays, which turn out to be small but
non-negligible. After factorization the long-distance QCD 
effects on $\bar B\to D_s^{(*)-} X_q$ decays are contained in two matrix 
elements: One is related to the decay constant for the $D_s^{(*)}$ meson,
accounting for the direct generation of the $D_s^{(*)}$ meson from the vacuum
by the $\bar c s$ current forming the weak Hamiltonian, while 
the other incorporates the effect of the initial $b$ quark bound state in the
$B$ meson, which relates a $b$-quark decay to a $B$-meson decay. The treatment
of the initial bound state effect will be the main
focus of this paper. We shall use the light-cone expansion and the heavy quark 
effective theory (HQET) \cite{hqet} to make our calculations aimed at 
understanding this effect from QCD.

\section{Factorization}
Both $\bar B\to D_s^{(*)-} X_c$ and $\bar B\to D_s^{(*)-} X_u$ involve tree 
amplitudes. In addition, penguin amplitudes contribute to 
$\bar B\to D_s^{(*)-} X_c$. The annihilation and exchange diagrams 
are estimated to give much smaller contributions 
than the tree diagram \cite{aleksan}. We shall neglect 
them. The relevant $\Delta B=1$ effective Hamiltonian reads

\begin{equation}
{\cal H}_{\rm eff} = {G_F\over \sqrt{2}}
\left\{ V_{qb} V^*_{cs}(c_1 O_1^q +c_2 O_2^q)-V_{tb}V^*_{ts}\sum_{n=3}^{10} c_n
O_n\right\},
\label{hamilton}
\end{equation}
where $O^q_{1,2}$ are the tree operators and $O_{3-6}$ $(O_{7-10})$ the
QCD (electroweak) penguin operators. They are given by

\begin{eqnarray}
&&O_1^q = (\bar q_i b_i)_{V-A} (\bar s_j c_j)_{V-A},\;\;
O_2^q = (\bar q_i b_j)_{V-A} (\bar s_j c_i)_{V-A},\nonumber\\
&&O_{3(5)} = (\bar s_i b_i)_{V-A}\sum_{q'}
(\bar q^\prime_j q^\prime_j)_{V-(+)A},\;\;
O_{4(6)} = (\bar s_i b_j)_{V-A}\sum_{q'}
(\bar q^\prime_j q^\prime_i)_{V-(+)A},\nonumber\\
&&O_{7(9)} = {3\over 2}(\bar s_i b_i)_{V-A}\sum_{q'}
e_{q^\prime}(\bar q^\prime_j q^\prime_j)_{V+(-)A},\;\;
O_{8(10)} ={3\over 2} (\bar s_i b_j)_{V-A}\sum_{q'}
e_{q^\prime}(\bar q^\prime_j q^\prime_i)_{V+(-)A},
\end{eqnarray}
where $i$ and $j$ are color indices, 
$(\bar q_1 q_2)_{V\pm A} = \bar q_1\gamma_\mu(1\pm\gamma_5)q_2$,
the summation runs over $q^\prime = u,d,s,c,b$, and $e_{q^\prime}$ is the 
electric charge of the $q^\prime$ quark in units of $|e|$.

The Wilson coefficients $c_n$ have been calculated in
different schemes \cite{BBL}. In this paper we shall consistently use the 
naive 
dimensional regularization scheme. Including the next-to-leading order QCD 
corrections, the values of $c_n$ at the renormalization scale $m_b$ 
are \cite{BBL}

\begin{eqnarray}
&&c_1 = 1.082,\;\;c_2 = -0.185,\;\;c_3=0.014,\;\;c_4 = -0.035,\;\;
c_5=0.009,\;\;c_6 = -0.041,\nonumber\\
&&c_7= -0.002\alpha_{em},\;\;c_8=0.054\alpha_{em},\;\;
c_9=-1.292\alpha_{em},\;\;c_{10}=0.263\alpha_{em}. \nonumber
\end{eqnarray}
Here $\alpha_{em}=1/137 $ is the electromagnetic fine structure constant.

Under the factorization hypothesis, we obtain from the effective Hamiltonian 
in Eq.~(\ref{hamilton}) the decay amplitudes

\begin{eqnarray}
&&A(\bar B\to D_s^- X_c) = if_{D_s}\left( \alpha p^\mu_{D_s}
\langle X_c|j^c_\mu|\bar B\rangle+\beta
\langle X_c|J|\bar B\rangle\right) , \label{ampli3}\\
&&A(\bar B\to D_s^{*-} X_c) = \alpha m_{D_s^*}f_{D_s^*}\epsilon^{\mu *}
\langle X_c|j^c_\mu|\bar B\rangle ,
\label{ampli4}\\
&&A(\bar B\to D_s^{-} X_u) = i{G_F\over \sqrt{2}}V_{ub} V^*_{cs} a_1
f_{D_s}p^\mu_{D_s}\langle X_u|j^u_\mu|\bar B\rangle , \label{ampli1}\\
&&A(\bar B\to D_s^{*-} X_u) = {G_F\over \sqrt{2}}V_{ub} V^*_{cs} a_1
m_{D_s^*}f_{D_s^*}\epsilon^{\mu *}
\langle X_u|j^u_\mu|\bar B\rangle , \label{ampli2}
\end{eqnarray}
where 

\begin{eqnarray}
&&\alpha={G_F\over\sqrt{2}}[V_{cb} V^*_{cs} a_1-V_{tb} V^*_{ts}(a_4+a_{10})],\\
&&\beta={G_F\over\sqrt{2}}V_{tb} V^*_{ts}(a_6+a_8){2m^2_{D_s}\over m_c+m_s},\\
&&a_{2i}=c_{2i}+{c_{2i-1}\over N_c},\\
&&a_{2i-1}=c_{2i-1}+{c_{2i}\over N_c}.
\end{eqnarray}
$N_c$ is the number of colors. The currents 
$j^q_\mu = \bar q\gamma_\mu(1-\gamma_5)b$ and $J=\bar c(1-\gamma_5)b$.

We have defined the decay constants of the pseudoscalar $D_s$ meson and
the vector $D^*_s$ meson as

\begin{eqnarray}
&&\langle D_s^-(p_{D_s})|\bar s\gamma^\mu\gamma^5 c|0\rangle = 
if_{D_s}p^\mu_{D_s}, \label{fd}\\
&&\langle D_s^{*-}(p_{D^*_s}, \epsilon)|\bar s\gamma_\mu c|0\rangle = 
m_{D^*_s}f_{D^*_s}\epsilon_\mu^{(\lambda)*}, \label{fdstar}
\end{eqnarray}
where $\epsilon$ stands for the polarization vector of the $D^*_s$ meson.
Using the definition (\ref{fd}) and the Dirac equation, it follows that

\begin{equation}
\langle D_s^-|\bar s(1+\gamma_5)c|0\rangle = if_{D_s}{m^2_{D_s}\over m_c+m_s},
\end{equation}
which has been used to obtain Eq.~(\ref{ampli3}).

\section{Initial bound state effect}
In this section we study the initial $b$-quark bound state effect on 
$\bar B\to D_s^{(*)-} X_q$ using the light-cone expansion and the heavy quark 
effective
theory. In the following we work out, in detail, the formulation for 
$\bar B\to D_s^- X_c$. The results for 
$\bar B\to D_s^{*-} X_c$ and $\bar B\to D_s^{(*)-} X_u$, which have a simpler 
structure in decay amplitudes, can be easily obtained along the same lines.

The differential decay rate for $\bar B\to D_s^- X_c$ in the $\bar B$ rest 
frame is given by

\begin{equation}
d\Gamma (\bar B\to D_s^- X_c)
= {1\over 2m_B} {d^3 {\bf p}_{D_s}\over (2\pi)^3 2 E_{D_s}}
\sum_{X_c} (2\pi)^4 \delta^4(p_B-p_{D_s}-p_{X_c}) |A(\bar B\to D_s^- X_c)|^2.
\label{rate}
\end{equation}
Applying $\int d^4y\,\, \mbox{exp}[-iy\cdot (p_B-p_{D_s}-p_{X_c})] = 
(2\pi)^4 \delta^4 (p_B-p_{D_s}-p_{X_c})$, 
$\langle X_c|Q_\mu(0)|\bar B\rangle = \langle X_c|Q_\mu(y)|\bar B\rangle
\mbox{exp}[-iy\cdot (p_{X_c}-p_B)]$ due to translation invariance, and the
completeness of the hadronic states $|X_c\rangle$,
we obtain from Eq.~(\ref{ampli3})

\begin{eqnarray}
\sum_{X_c}&&(2\pi)^4 \delta^4(p_B-p_{D_s}-p_{X_c}) |A(\bar B\to D_s^- X_c)|^2
\nonumber\\
&&= f_{D_s}^2\int d^4y e^{iy\cdot p_{D_s}}
( |\alpha|^2 p^\mu_{D_s}p^\nu_{D_s}
\langle\bar B|[j_\nu^{c\dagger}(0),j_\mu^c(y)]|\bar B\rangle 
+|\beta|^2\langle\bar B|[J^\dagger(0),J(y)]|\bar B\rangle \nonumber\\
&&+\alpha\beta^* p_{D_s}^\mu
\langle\bar B|[J^\dagger(0),j_\mu^c(y)]|\bar B\rangle+
\alpha^*\beta p_{D_s}^\mu
\langle\bar B|[j_\mu^{c\dagger}(0),J(y)]|\bar B\rangle ). 
\label{square1}
\end{eqnarray}

Computing the commutators of currents yields

\begin{eqnarray}
\sum_{X_c}&&(2\pi)^4 \delta^4(p_B-p_{D_s}-p_{X_c}) |A(\bar B\to D_s^- X_c)|^2
\nonumber\\
&&= -2f_{D_s}^2\left( |\alpha|^2 S_{\mu\alpha\nu\beta}
p^\mu_{D_s}p^\nu_{D_s} + |\beta|^2 g_{\alpha\beta}\right) 
\int d^4y e^{iy\cdot p_{D_s}}\left[ \partial^{\alpha}\Delta_c (y) \right] 
\langle\bar B|\bar{b}(0)\gamma^{\beta}U(0, y)b(y)|\bar B\rangle \nonumber\\
&&-2f_{D_s}^2(\alpha\beta^*+\alpha^*\beta)m_c p_{D_s}^\beta
\int d^4y e^{iy\cdot p_{D_s}} i\Delta_c (y) 
\langle\bar B|\bar{b}(0)\gamma_\beta U(0, y)b(y)|\bar B\rangle ,
\label{square2}
\end{eqnarray}
where $S_{\mu\alpha\nu\beta} = g_{\mu\alpha}g_{\nu\beta} + g_{\mu\beta}
 g_{\nu\alpha} - g_{\mu\nu}g_{\alpha\beta}$ and

\begin{equation}
\Delta_q(y) = -{i\over (2\pi)^3} \int d^4k\,\, e^{-ik\cdot y}\varepsilon(k^0)
\delta(k^2-m_q^2).
\end{equation}
The soft gluon interactions on the final state $c$ quark appear in
the Wilson line operator

\begin{equation}
U(x,y) = {\cal P} \mbox{exp}[ig_s \int^x_y dz^\mu A_\mu(z)],
\end{equation}
where ${\cal P}$ denotes path-ordering and $A^\mu$ is the background gluon 
field. The initial bound state effect is contained
in the matrix element of the non-local $b$-quark operator,
$\langle\bar B|\bar{b}(0)\gamma^{\beta}U(0, y)b(y)|\bar B\rangle$.
We note that the same matrix element also incorporates bound state effects
in other inclusive $B$-meson decays, including
$B\to \ell\bar\nu_\ell X_q$ \cite{jp}, $B\to \gamma X_s$ \cite{rare}, and 
$B\to K^{(*)}X$ \cite{KX}. 
We shall use the light-cone expansion to calculate the matrix element. 
The light-cone expansion enables a systematic ordering of nonperturbative
QCD effects as an expansion in powers of a small parameter of order 
$\Lambda^2_{\rm QCD}/m^2_B$.
The general method has been described in \cite{jp,rare}.
We shall restrict ourselves here to the essential steps only.

Each integrand of the integrals in Eq.~(\ref{square2}) contains an oscillating 
factor 
$e^{iy\cdot p_{D_s}}$. The dominant contribution comes from domains with less 
rapid oscillations, i.e., $|y\cdot p_{D_s}|\sim 1$. This implies that for 
sufficiently large momentum of the outgoing $D_s$ meson, 
the decay dynamics is
dominated by spacetime separations in the neighborhood of the light cone
$y^2 = 0$. The light-cone dominance is also justified since
the function $\Delta_c(y)$ in the integrals in Eq.~(\ref{square2}) 
has a singularity at $y^2 = 0$. Therefore, at leading twist 
we have\footnote{Higher twist effects have been analyzed quantitatively
in \cite{HT}. They give rise to corrections of order 
$\Lambda_{\rm QCD}^2/m^2_B$ and can be added systematically. The calculation
of higher twist effects on the processes under consideration is 
beyond the scope of this paper.}

\begin{equation}
\langle\bar B|\bar{b}(0)\gamma^{\beta}U(0, y)b(y)|\bar B\rangle =
2p_B^\beta\int_0^1 d\xi e^{-i\xi y\cdot p_B} f(\xi),
\end{equation}
where $f(\xi)$ is the $b$-quark distribution function of the $B$ 
meson introduced in Refs.~\cite{jp,rare}

\begin{equation}
f(\xi)= \frac{1}{4\pi}\int\frac{d(y\cdot p_B)}{y\cdot p_B}e^{i\xi y\cdot p_B}
\langle\bar B|\bar b(0)y\!\!\!/ U(0, y)b(y)|\bar B\rangle |_{y^2=0} .
\label{eq:def}
\end{equation} 
Note that $f(\xi)$ depends only on the properties of the $B$-meson bound
state and is not specific to particular $B$-meson decay process in question. 
Thus, $f(\xi)$ is a universal distribution as fundamental as parton 
distributions in
deep inelastic scattering. It has the interpretation of the probability of
finding a $b$-quark with momentum $\xi p_B$ inside the $B$ meson with momentum
$p_B$ \cite{rare,parton}.

Assembling all the pieces, 
we obtain the momentum spectrum for $D_s^-$ in $\bar B\to D_s^- X_c$

\begin{eqnarray}
{d\Gamma\over d|{\bf p}_{D_s}|}(\bar B\to D_s^- X_c) =
&&{f^2_{D_s}\over \pi}
{|{\bf p}_{D_s}|^2\over E_{D_s}}
\int_0^1 d\xi f(\xi)\varepsilon (E_{D_s}-m_B\xi) \nonumber\\
&&\times\delta (m_B^2\xi^2 -
2m_B E_{D_s}\xi + m_{D_s}^2-m_c^2) \nonumber\\
&&\times [|\alpha|^2(m^2_{D_s}E_{D_s}+m_B m^2_{D_s}\xi-2m_B E_{D_s}^2\xi) 
+|\beta|^2(E_{D_s}-m_B\xi) \nonumber\\
&&-(\alpha\beta^*+\alpha^*\beta)m_cE_{D_s}].
\label{dspectrum}
\end{eqnarray}
Carrying out the $\delta$-function integration in Eq.~(\ref{dspectrum}), we 
finally arrive at

\begin{eqnarray}
{d\Gamma\over d|{\bf p}_{D_s}|}(\bar B\to D_s^- X_c) =
&&{f^2_{D_s}\over 2\pi m_B}
{|{\bf p}_{D_s}|^2\over E_{D_s}\sqrt{|{\bf p}_{D_s}|^2+m_c^2}}\nonumber\\
&&\times\{ f(\xi_+^c) [ |\alpha|^2 (2m_BE_{D_s}^2\xi_+^c-
m_Bm^2_{D_s}\xi_+^c-m^2_{D_s}E_{D_s} ) \nonumber\\
&&+|\beta|^2 (m_B\xi_+^c-E_{D_s})+(\alpha\beta^*+\alpha^*\beta)m_cE_{D_s}]
\nonumber\\
&&-(\xi_+^c\to \xi_-^c ) \} ,
\label{spectrum1}
\end{eqnarray}
where

\begin{equation}
\xi_\pm^q = {E_{D_s}\pm\sqrt{|{\bf p}_{D_s}|^2+m^2_q}\over m_B}.
\end{equation}
The interference between the tree and penguin amplitudes is associated with
the coefficients $|\alpha|^2$ and $\alpha\beta^*+\alpha^*\beta$.

In the case of the $\bar B\to D_s^{*-} X_c$ and $\bar B\to D_s^{(*)-} X_u$ 
decays, similar considerations lead to the momentum spectra in the $\bar B$ 
rest frame

\begin{eqnarray}
{d\Gamma\over d|{\bf p}_{D^*_s}|}(\bar B\to D_s^{*-} X_c) =
&&{f^2_{D^*_s}\over 2\pi m_B}|\alpha|^2
{|{\bf p}_{D^*_s}|^2\over E_{D^*_s}\sqrt{|{\bf p}_{D^*_s}|^2+m_c^2}}\nonumber\\
&&\times\left[ f(\xi_+^{*c})\left( 2m_BE_{D^*_s}^2\xi_+^{*c}+
m_Bm^2_{D^*_s}\xi_+^{*c}-3m^2_{D^*_s}E_{D^*_s}\right)-
\left( \xi_+^{*c}\to \xi_-^{*c}\right) \right] ,
\label{spectrum2}\\
{d\Gamma\over d|{\bf p}_{D_s}|} (\bar B\to D_s^{-} X_u) =
&&{G_F^2\over 4\pi m_B}|V_{ub} V^*_{cs}|^2 a_1^2 f^2_{D_s}
{|{\bf p}_{D_s}|^2\over E_{D_s}\sqrt{|{\bf p}_{D_s}|^2+m_u^2}}\nonumber\\
&&\times\left [f(\xi_+^{u}) \left (2m_BE_{D_s}^2\xi_+^{u} -
m_Bm^2_{D_s}\xi_+^{u} 
-m^2_{D_s}E_{D_s}\right )-\left(\xi_+^{u}\to \xi_-^{u}\right)\right ],
\label{spectrum3}\\
{d\Gamma\over d|{\bf p}_{D^*_s}|} (\bar B\to D_s^{*-} X_u) =
&&{G_F^2\over 4\pi m_B}|V_{ub} V^*_{cs}|^2 a_1^2 f^2_{D^*_s}
{|{\bf p}_{D^*_s}|^2\over E_{D^*_s}\sqrt{|{\bf p}_{D^*_s}|^2+m_u^2}}\nonumber\\
&&\times\left [f(\xi_+^{*u}) \left (2m_BE_{D^*_s}^2\xi_+^{*u} +
m_Bm^2_{D^*_s}\xi_+^{*u} 
-3m^2_{D^*_s}E_{D^*_s}\right )-\left(\xi_+^{*u}\to \xi_-^{*u}\right)\right ],
\label{spectrum4}
\end{eqnarray}
where

\begin{equation}
\xi_{\pm}^{*q} = {E_{D^*_s}\pm\sqrt{|{\bf p}_{D^*_s}|^2+m^2_q}\over m_B}.
\end{equation}

For a longitudinally polarized $D_s^{*-}$ we obtain

\begin{eqnarray}
{d\Gamma_L\over d|{\bf p}_{D^*_s}|}&&[\bar B\to D_s^{*-}(\lambda=0) X_c] 
\nonumber\\
&&={f^2_{D^*_s}\over 2\pi m_B}|\alpha|^2
{|{\bf p}_{D^*_s}|^2\over E_{D^*_s}\sqrt{|{\bf p}_{D^*_s}|^2+m_c^2}}\nonumber\\
&&\times\left[ f(\xi_+^{*c})\left( 2m_B|{\bf p}_{D^*_s}|^2\xi_+^{*c}+
m_Bm^2_{D^*_s}\xi_+^{*c}-m^2_{D^*_s}E_{D^*_s}\right)
-\left( \xi_+^{*c}\to \xi_-^{*c}\right) \right] .
\label{spectrum5}
\end{eqnarray}

For a transversely polarized $D_s^{*-}$ we have

\begin{eqnarray}
{d\Gamma_T\over d|{\bf p}_{D^*_s}|}&&[\bar B\to D_s^{*-}(\lambda=\pm 1) X_c]
\nonumber\\
&&={m^2_{D^*_s}f^2_{D^*_s}\over \pi m_B}|\alpha|^2
{|{\bf p}_{D^*_s}|^2\over E_{D^*_s}\sqrt{|{\bf p}_{D^*_s}|^2+m_c^2}}
\left[ f(\xi_+^{*c})\left( m_B\xi_+^{*c}-E_{D^*_s}\right)-
\left( \xi_+^{*c}\to \xi_-^{*c}\right) \right] .
\label{spectrum6}
\end{eqnarray}

From Eq.~(\ref{spectrum1}) one can easily obtain the hadronic invariant mass 
spectrum for $X_c$ in $\bar B\to D_s^- X_c$ using the relation

\begin{equation}
{d\Gamma\over dm_{X_c}}(\bar B\to D_s^- X_c) =
{m_{X_c}E_{D_s}\over m_B|{\bf p}_{D_s}|}
{d\Gamma\over d|{\bf p}_{D_s}|}(\bar B\to D_s^- X_c).
\end{equation}
Similar formulas hold for the hadronic invariant mass spectra in  
$\bar B\to D_s^{*-} X_c$ and $\bar B\to D_s^{(*)-} X_u$.

The total decay rates can be obtained by integrating Eqs.~(\ref{spectrum1}),
(\ref{spectrum2})-(\ref{spectrum4}), (\ref{spectrum5}) and (\ref{spectrum6})
with respect to $|{\bf p}_{D_s^{(*)}}|$ over the entire kinematic range

\begin{equation}
0\leq |{\bf p}_{D_s^{(*)}}|\leq 
\sqrt{[(m_B+m_{D_s^{(*)}})^2-m^2_{X_{q_{\rm min}}}]
[(m_B-m_{D_s^{(*)}})^2-m^2_{X_{q_{\rm min}}}]}/(2m_B),
\label{range}
\end{equation}
where $m_{X_{q_{\rm min}}}$ is the minimum value of the invariant mass of the
hadronic system $X_q$.

Measurements of the ratio of the decay rate for the longitudinally polarized
$D_s^{*-}$ to the total decay rate for $\bar B\to D_s^{*-}X_c$ have a good
potential for testing the theory, since the theoretical uncertainties partly
cancel in the ratio and most of experimental systematic errors also cancel.

The leading effect of the initial $B$ bound state is encoded in the 
distribution function $f(\xi)$. Since it is a nonperturbative QCD object,
$f(\xi)$ has not yet been completely determined. Nevertheless, several
important properties of it are known in QCD \cite{jp,rare}. The distribution 
function is
exactly normalized to unity, $\int_0^1d\xi f(\xi)=1$, because of the 
conservation of the $b$-quark vector current by the strong interactions.
For a free $b$-quark, $b(y)=e^{-iy\cdot p_b}b(0)$. From Eq.~(\ref{eq:def}) it
follows then that in the free quark limit

\begin{equation}
f_{\rm free}(\xi)=\delta\left(\xi-{m_b\over m_B}\right).
\label{freef}
\end{equation}
The mean $\langle\xi\rangle = \int^1_0 d\xi \xi f(\xi)$ and
the variance $\sigma^2 = \langle(\xi-\langle\xi\rangle)^2\rangle =
\int^1_0 d\xi (\xi-\langle\xi\rangle)^2 f(\xi)$ 
characterize the 
location of the ``center of mass'' of the distribution function and the square
of its width, respectively. They specify the gross shape of the 
distribution function. They have been estimated using
the heavy quark effective theory. The results are \cite{jp,rare}

\begin{eqnarray}
&&\langle\xi\rangle = {m_b\over m_B}\left[1-{5\over 6m^2_b}(\lambda_1+
3\lambda_2)\right],
\label{mean}\\
&&\sigma^2 = -{\lambda_1\over 3m_B^2},
\label{width}
\end{eqnarray}
where $\lambda_1$ and $\lambda_2$ are the HQET parameters, which are defined as

\begin{eqnarray}
\lambda_1 &=& \frac{1}{2m_B}\langle \bar B|\bar h_v
(iD)^2 h_v|\bar B\rangle , \label{eq:mK} \\
\lambda_2 &=& \frac{1}{12m_B}\langle\bar B|\bar h_v g_sG_{\mu\nu}
\sigma^{\mu\nu} h_v|\bar B\rangle . 
\label{eq:mG}
\end{eqnarray}
Both of them are expected to be of order $\Lambda^2_{\rm QCD}$.
These imply that the distribution function is sharply peaked around
$m_b/m_B$ with a width of order $\Lambda_{\rm QCD}/m_B$. This is in line with 
the expectation that the $b$ quark should carry most of the momentum of the 
$B$ meson.

Another important property of the distribution function, discussed previously,
is universality. This implies that the distribution function can be extracted
from measurements of one process, and then applied to predict 
others in a model-independent way. Indeed,
the shape of the distribution function can be directly extracted
from $B\to \ell\bar\nu_\ell X_q$ \cite{shape} or $B\to \gamma X_s$ \cite{rare} 
decays. This could eventually improve the precision of theoretical predictions.

For numerical analyses we take the following parameterization for the 
distribution function \cite{jin3}

\begin{eqnarray}
f(\xi) = N {\xi (1-\xi)^c\over [(\xi -a)^2 +b^2]^d},
\label{dis}
\end{eqnarray}
where $N$ is a normalization constant which guarantees
$\int^1_0 d \xi f(\xi) = 1$. The four shape parameters, $(a,b,c,d)$, obey 
all the known constraints on the distribution function.
Once the parameters $c$ and $d$ are given, the parameters $a$ and $b$ can be
fixed by comparing with $\langle\xi\rangle$ and $\sigma^2$ given in 
Eqs.~(\ref{mean}) and
(\ref{width}). Unfortunately, we do not know 
the values for $c$ and $d$ at present. We shall take $c$ and $d$
to be free parameters and vary them to gain an idea of how the momentum 
spectra for $D_s^{(*)}$ and branching fractions depend on the shape of 
$f(\xi)$. Eventually, the shape of the distribution function can be determined
from experimental data. The parameterization (\ref{dis}) facilitates the 
comparison to data.

\section{Free quark limit}
In order to understand the bound state dynamics of the $B$ meson in the 
semi-inclusive $B$ meson decays, we should consider the 
decay rates for the free quark decays, $b\to D_s^{(*)-} q$, which give us a
reference point.

In the free quark limit, the distribution function becomes a $\delta$-function.
Substituting the free $b$-quark distribution function Eq.~(\ref{freef}) in 
Eqs.~(\ref{spectrum1}) [or, most easily, Eq.~(\ref{dspectrum})] and 
(\ref{spectrum2})-(\ref{spectrum4}), we reproduce the corresponding total
decay rates for the free quark decays in the rest frame of the $b$ quark:

\begin{eqnarray}
\Gamma(b\to D_s^{-}c)=&&{f^2_{D_s}\over 4\pi m_b^2}|{\bf p}_{D_s}|
\{ |\alpha|^2 [(m_b^2-m_c^2)^2-m_{D_s}^2(m_b^2+m_c^2)]+
|\beta|^2 (m_b^2+m_c^2-m^2_{D_s}) \nonumber\\
&&+(\alpha\beta^*+\alpha^*\beta)m_c(m_b^2-m_c^2+m_{D_s}^2)\}, 
\label{freewid1}
\end{eqnarray}
\begin{eqnarray}
&&\Gamma(b\to D_s^{*-}c)={f^2_{D^*_s}\over 4\pi m_b^2} |\alpha|^2 
|{\bf p}_{D^*_s}| [(m_b^2-m_c^2)^2+m_{D^*_s}^2 (m_b^2+m_c^2-2m^2_{D^*_s})], 
\label{freewid2}\\
&&\Gamma(b\to D_s^{-}u)={G_F^2\over 8\pi m_b^2}|V_{ub} V^*_{cs}|^2 a_1^2 
f^2_{D_s} |{\bf p}_{D_s}|
[(m_b^2-m_u^2)^2-m_{D_s}^2 (m_b^2+m_u^2)], \label{freewid3}\\
&&\Gamma(b\to D_s^{*-}u)={G_F^2\over 8\pi m_b^2}|V_{ub} V^*_{cs}|^2 a_1^2 
f^2_{D^*_s}|{\bf p}_{D^*_s}|
[(m_b^2-m_u^2)^2+m_{D^*_s}^2(m_b^2+m_u^2-2m_{D^*_s}^2)]. \label{freewid4}
\end{eqnarray}

Similarly, substituting Eq.~(\ref{freef}) in Eqs.~(\ref{spectrum5}) and 
(\ref{spectrum6}), we reproduce the corresponding decay rates for 
$b\to D_s^{*-} c$ with the 
longitudinal and transverse polarizations of $D_s^{*-}$, respectively, 

\begin{eqnarray}
&&\Gamma_L[b\to D_s^{*-}(\lambda=0)c]={f^2_{D^*_s}\over 4\pi m_b^2} 
|\alpha|^2 |{\bf p}_{D^*_s}| [(m_b^2-m_c^2)^2-m_{D^*_s}^2 (m_b^2+m_c^2) ], 
\label{freewid5}\\
&&\Gamma_T[b\to D_s^{*-}(\lambda=\pm 1)c]=
{m^2_{D^*_s}f^2_{D^*_s}\over 2\pi m_b^2}
|\alpha|^2 |{\bf p}_{D^*_s}| (m_b^2+m_c^2-m_{D^*_s}^2). 
\label{freewid6}
\end{eqnarray}

It follows from Eqs.~(\ref{freewid2}) and (\ref{freewid5}) that the ratio of
the decay rate for the longitudinally polarized
$D_s^{*-}$ to the total decay rate for $b\to D_s^{*-}c$ 

\begin{equation}
{\Gamma_L\over \Gamma}(b\to D_s^{*-}c)= 1-
{2m^2_{D^*_s}(m_b^2+m_c^2-m^2_{D_s^*})\over (m_b^2-m_c^2)^2+m^2_{D_s^*}
(m_b^2+m_c^2-2m^2_{D_s^*})}.
\label{freeratio}
\end{equation}

Our results for the part of the free quark decay rate involving tree diagrams
agree with Refs.~\cite{palmer,aleksan}.
It should be stressed that the reproduction of the free quark decay results by
taking the free quark limit of the leading twist results shows consistency
of the light-cone expansion.
  
All the momentum spectra for $D_s^{(*)-}$ in the free quark decays 
$b\to D_s^{(*)-} q$ are a discrete line localized at 

\begin{equation}
|{\bf p}_{D^{(*)}_s}|=
\sqrt{[(m_b+m_{D^{(*)}_s})^2-m_q^2][(m_b-m_{D^{(*)}_s})^2-m_q^2]}/(2m_b)
\label{line}
\end{equation}
because the momentum of $D_s^{(*)-}$ in the two-body decay is fixed 
kinematically.

As shown in the last section, at the hadron level the momentum spectrum for
$D_s^{(*)-}$ in $\bar B\to D_s^{(*)-}X_q$ spreads over the kinematic range 
given in Eq.~(\ref{range}), due to the fact that the hadronic invariant mass 
$m_{X_q}$ is changeable. However, since the 
heavy $b$ quark in the $B$ meson is nearly free, most of the spectrum remain 
around the free quark decay location of (\ref{line}). This is the phase space 
region where light-cone dominance occurs. 

For measurements of 
$\bar B\to D_s^{(*)-}X_u$ one can apply the cut on the $D_s^{(*)-}$ momentum
above the limit for charm production in the decay, 
$|{\bf p}_{D_s^{(*)}}| > 
\sqrt{[(m_B+m_{D_s^{(*)}})^2-m^2_D]
[(m_B-m_{D_s^{(*)}})^2-m^2_D]}/(2m_B)$,
to suppress the $\bar B\to D_s^{(*)-}X_c$ background. Since this cut is well
below the free quark decay location of (\ref{line}), the majority of the
momentum spectrum in $\bar B\to D_s^{(*)-}X_u$ pass the cut, as we shall see.

\section{Numerical analyses}
For numerical analyses, we must specify the values of the input parameters.
We use $G_F=1.16639\times 10^{-5}$ GeV$^{-2}$, $m_B=5.279$
GeV, $m_{D_s}=1.969$ GeV, $m_{D^*_s}=2.112$ GeV, $m_D=1.87$ GeV, and $m_\pi=0$. We 
neglect the $B^+$, $B^0$ lifetime difference and use 
$\tau_B=1.6\times 10^{-12}$ sec. For quark
masses, we use $m_b=4.9$ GeV,
$m_c=1.5$ GeV, $m_s=120$ MeV, and $m_u=0$. As for the CKM matrix elements, we
take $|V_{ub}|=0.0035$, $|V_{cs}|=0.9742$, $|V_{cb}|=|V_{ts}|=0.04$, and
$|V_{tb}|=0.9992$. Since the phase 
$\omega_{ud}\equiv {\rm arg}(-V^*_{cs}V^*_{tb}V_{cb}V_{ts})$ is very small, we 
assume $\cos\omega_{ud}=1$. For the decay constants, we take $f_{D_s}=264$
MeV from the measurements of the leptonic branching fractions
Br($D_s\to\tau\nu$) and Br($D_s\to\mu\nu$) \cite{fds} and assume
$f_{D^*_s}=f_{D_s}$.  We take $m_{X_{c_{\rm min}}}=m_c$
for $\bar B\to D_s^{(*)-}X_c$ and $m_{X_{u_{\rm min}}}=0$
for $\bar B\to D_s^{(*)-}X_u$.
Finally, we consider two very different shapes of the
distribution function parameterized in Eq.~(\ref{dis}), corresponding to two
sets of parameters:
(i) preset $c=d=1$, in that case $a=0.9548$ and $b=0.005444$, which are 
inferred from
the known mean value and variance of the distribution function given by
Eqs.~(\ref{mean}) and (\ref{width}) using the HQET parameters 
$\lambda_1= -0.5$ GeV$^2$ and $\lambda_2= 0.12$ GeV$^2$; (ii) preset
$c=d=2$, in that case $a=0.9864$ and $b=0.02557$ inferred from the same mean
value and variance of the distribution function \cite{jin3}.

We first compute the branching fractions for $\bar B\to D_s^{(*)-}X_q$. It 
turns out that the branching fractions are not sensitive to the detailed shape
of the distribution function. The difference between the branching fractions
obtained using the parameter sets (i) and (ii) for the distribution function is 
negligible. 
The results for the direct production of $D_s^{(*)-}$ in 
semi-inclusive B decays are

\begin{eqnarray} 
&&{\cal B}(\bar B\to D_s^{-}X_c)=4.0\%, \nonumber\\
&&{\cal B}(\bar B\to D_s^{*-}X_c)=5.4\%,\nonumber\\ 
&&{\cal B}(\bar B\to D_s^{-}X_u)=4.8\times 10^{-4}, \nonumber\\
&&{\cal B}(\bar B\to D_s^{*-}X_u)=6.2\times 10^{-4}.
\label{predic}
\end{eqnarray}  
The calculated branching fraction for $\bar B\to D_s^{*-}X_c$ is in agreement
with the measured one by BABAR quoted in the Introduction.
Because $D_s^{*-}$ decays to $D_s^-\gamma$ or $D_s^-\pi^0$, we can
compare the sum of both channels, 
${\cal B}(\bar B\to D_s^{-}X_c)+{\cal B}(\bar B\to D_s^{*-}X_c)=9.4\%$,
from Eq.~(\ref{predic}) with the measured branching fraction for $B\to D_s^{\pm}X$ 
quoted in the Introduction. The comparison shows an encouraging agreement.
 
We find that the contribution of the penguin amplitude decreases
the branching fraction for $\bar B\to D_s^{-}X_c$ by $3\%$ and the branching
fraction for $\bar B\to D_s^{*-}X_c$ by $6\%$. So the penguin contribution is
small but non-negligible, especially for $\bar B\to D_s^{*-}X_c$. 

Comparing with the calculations in the free quark decays $b\to D_s^{(*)-}q$,
we find that all the branching fractions for $\bar B\to D_s^{(*)-}X_q$ are 
enhanced by the initial bound state effect by about $5\%$.

The ratio of the decay rate for the longitudinally polarized $D_s^{*-}$ to the
total decay rate for $\bar B\to D_s^{*-}X_c$ is also insensitive to the 
detailed shape of the distribution function. The bound state effect increases
the ratio by $2\%$ compared to the free quark decay. We obtain

\begin{equation}
{\Gamma_L\over\Gamma} (\bar B\to D_s^{*-}X_c) = 0.66.
\end{equation}

Now we turn to the differential decay rates. Figures 1-4 show the momentum 
spectra for $D_s^{(*)-}$ in $\bar B\to D_s^{(*)-}X_q$. We choose to use two
very different shapes of the distribution function to calculate the spectra.  
In contrast to the integrated decay rates, all the spectra depend
strongly on the shape of the distribution function. Hence measurements of 
the spectra can provide useful information about the nonperturbative 
distribution function. 

As Figs.~3 and 4 make clear, the majority of the
momentum spectrum for $D_s^{(*)-}$ in the decay mode
$\bar B\to D_s^{(*)-}X_u$ lie above the kinematic limit for the dominated decay
mode $\bar B\to D_s^{(*)-}X_c$, which is $|{\bf p}_{D_s}| > 1.81$ GeV for
$\bar B\to D_s^{-}X_u$ and $|{\bf p}_{D^*_s}| > 1.73$ GeV for
$\bar B\to D_s^{*-}X_u$. The fraction of spectrum above the charmed limit is
not sensitive to the shape of the distribution function. We find that about
$98\%$ of the spectrum goes beyond the charmed limit. Thus, the kinematic cut on
the momentum of $D_s^{(*)}$ provides an efficient discrimination between 
$\bar B\to D_s^{(*)-}X_u$ decays and $\bar B\to D_s^{(*)-}X_c$ decays. 

\section{Discussion and Conclusion} 
We have calculated the total and differential $\bar B\to D_s^{(*)-}X_q$ decay
rates assuming factorization. We have treated the initial
bound state effect using the light-cone expansion and the heavy quark effective
theory, generating a theoretical description of this effect from QCD  rather than 
the phenomenological
model \cite{aleksan}. The resulting branching fractions for $\bar B\to D_s^{(*)-}X_c$ are
found to be in agreement with the measurements. This suggests that factorization 
is a fair approximation for these decay modes. However,
there are still considerable uncertainties in the theoretical calculations and
experimental measurements. Theoretical uncertainties arise from the input parameters,
the $b$-quark distribution function, the higher twist corrections, and the 
radiative corrections. The uncertainty in the decay rate from the error on the 
decay constant $f_{D_s}$ alone \cite{fds} is already quite large ($\sim 30\%$).

For the determination of $|V_{ub}|$ from
$\bar B\to D_s^{(*)-}X_u$, additional theoretical uncertainty results from the
underlying factorization hypothesis itself. It is hard to quantify this uncertainty. 
Therefore, $|V_{ub}|$ determined from the semi-inclusive 
nonleptonic decay $\bar B\to D_s^{(*)-}X_u$ will
not be competitive with that determined from the inclusive semileptonic decay
$\bar B\to \ell\bar\nu_\ell X_u$ \cite{win02}. Nevertheless, $\bar B\to D_s^{(*)-}X_u$ will
provide a useful independent measurement of $|V_{ub}|$ with very different systematic
errors. On the other hand, using
$|V_{ub}|$ determined from the inclusive semileptonic $B$ decay, the range of 
validity of calculational tools for addressing QCD can be tested and established in
$\bar B\to D_s^{(*)-}X_u$,
 
\acknowledgments
It is a pleasure to thank Berthold Stech for useful discussions.
This work is supported by the Australian Research Council.

{\tighten
} 

\newpage
\begin{figure}[t]
\centerline{\epsfysize=9truecm \epsfbox{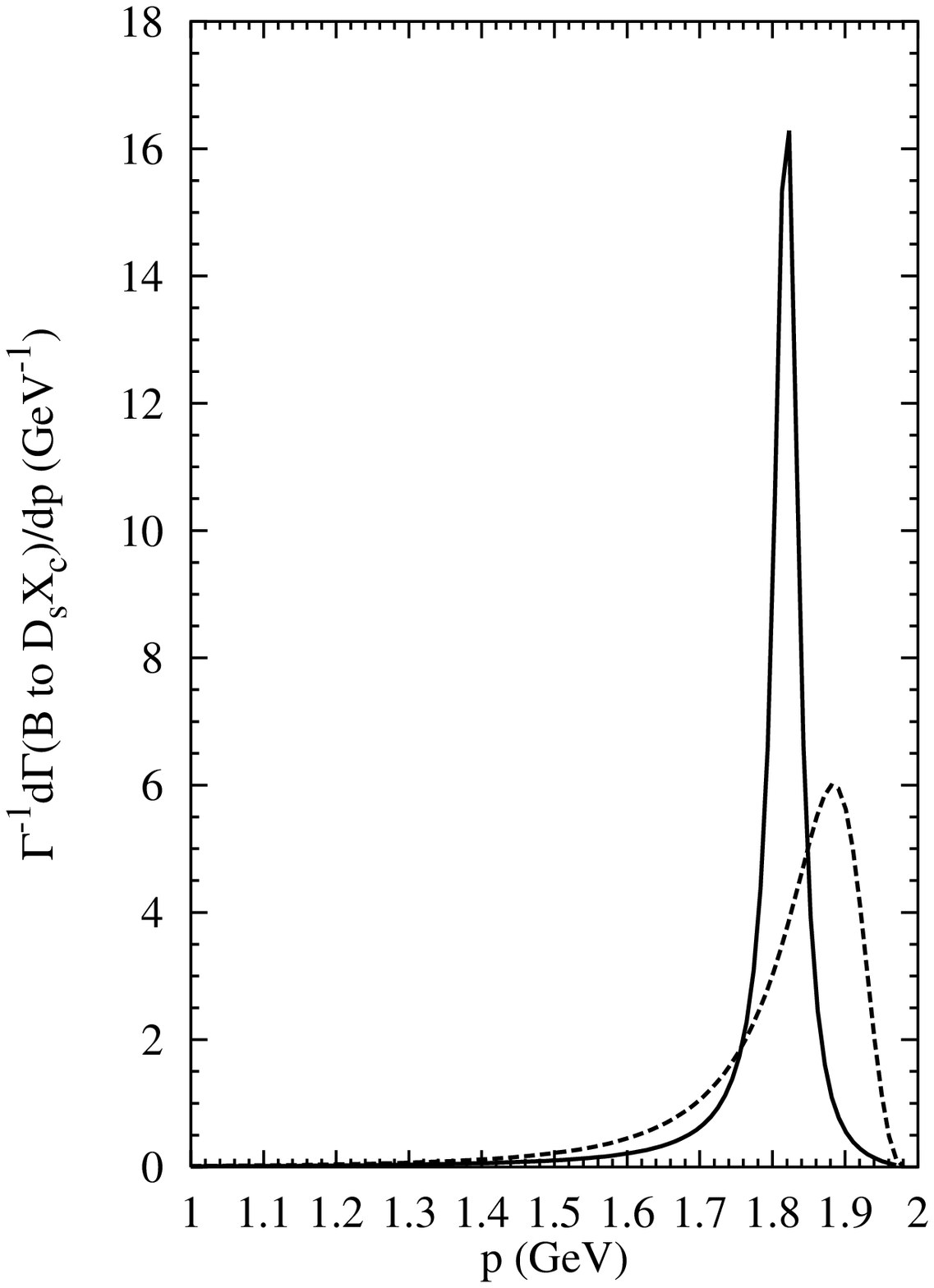}}
{\tighten
\caption {
Momentum spectrum for $D_s^-$ in $\bar B \to D_s^-  X_c$.
In Figs.~1-4, the solid curves are obtained using the parameter set (i) 
for the distribution function; the
dashed curves are obtained using the parameter set (ii) 
for the distribution function.}} 
\label{bc}
\end{figure}

\begin{figure}[ht]
\centerline{\epsfysize=9truecm \epsfbox{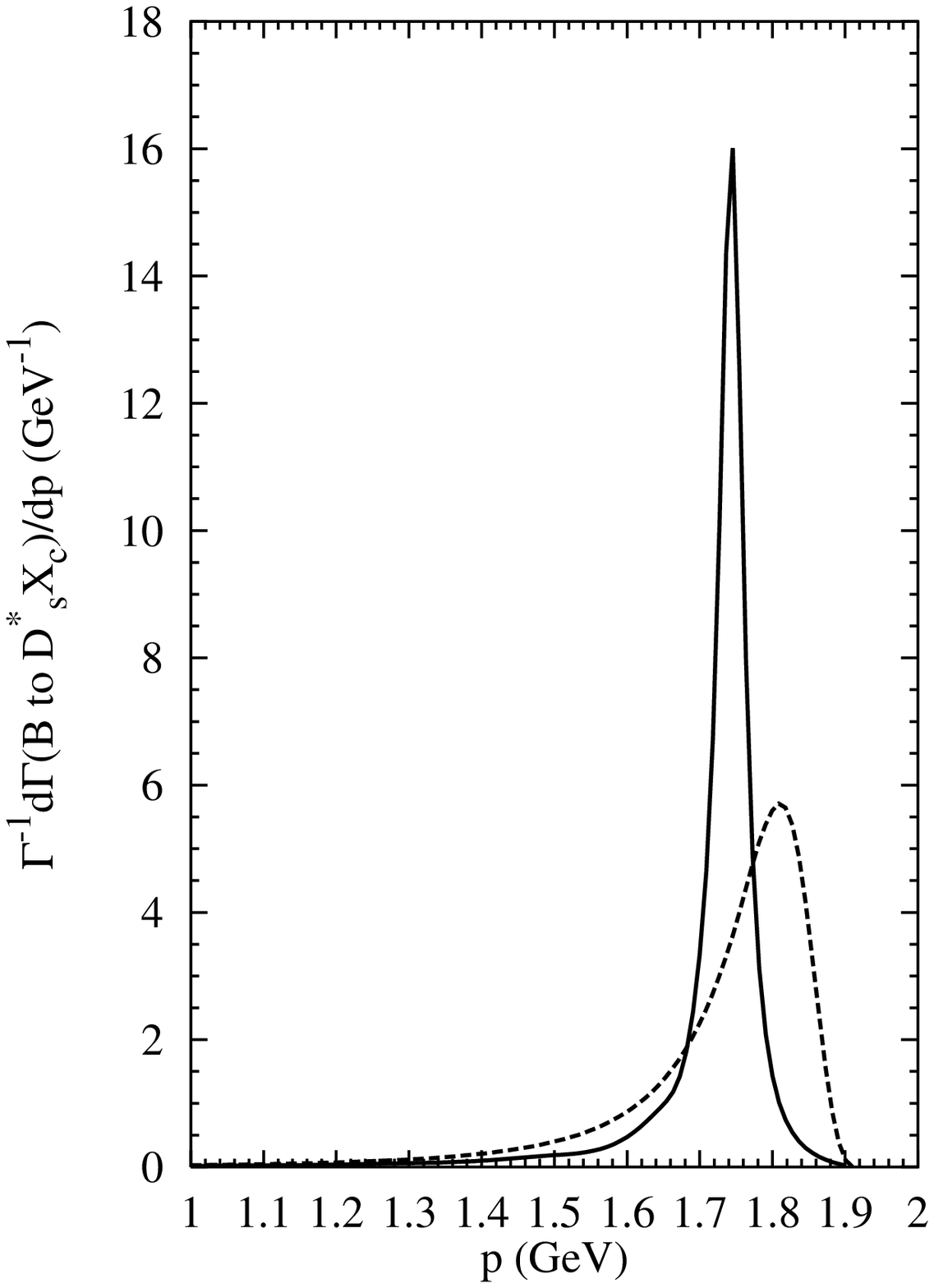}}
{\tighten
\caption {Momentum spectrum for $D_s^{*-}$ in $\bar B \to D_s^{*-} X_c$.
} }
\label{bcstar}
\end{figure}

\begin{figure}[t]
\centerline{\epsfysize=9truecm \epsfbox{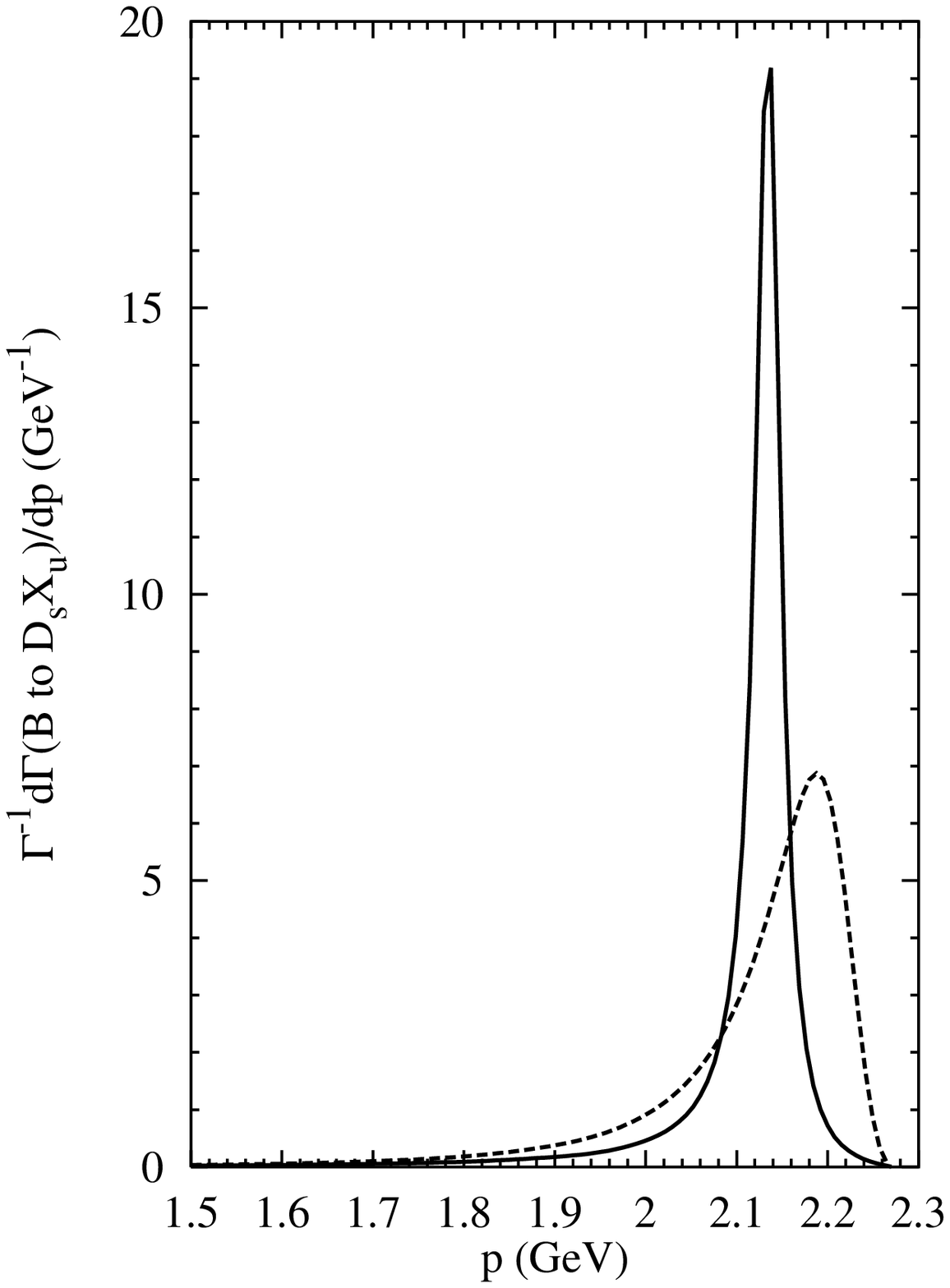}}
{\tighten
\caption {
Momentum spectrum for $D_s^-$ in $\bar B \to D_s^-  X_u$.
}} 
\label{bu}
\end{figure}

\begin{figure}[ht]
\centerline{\epsfysize=9truecm \epsfbox{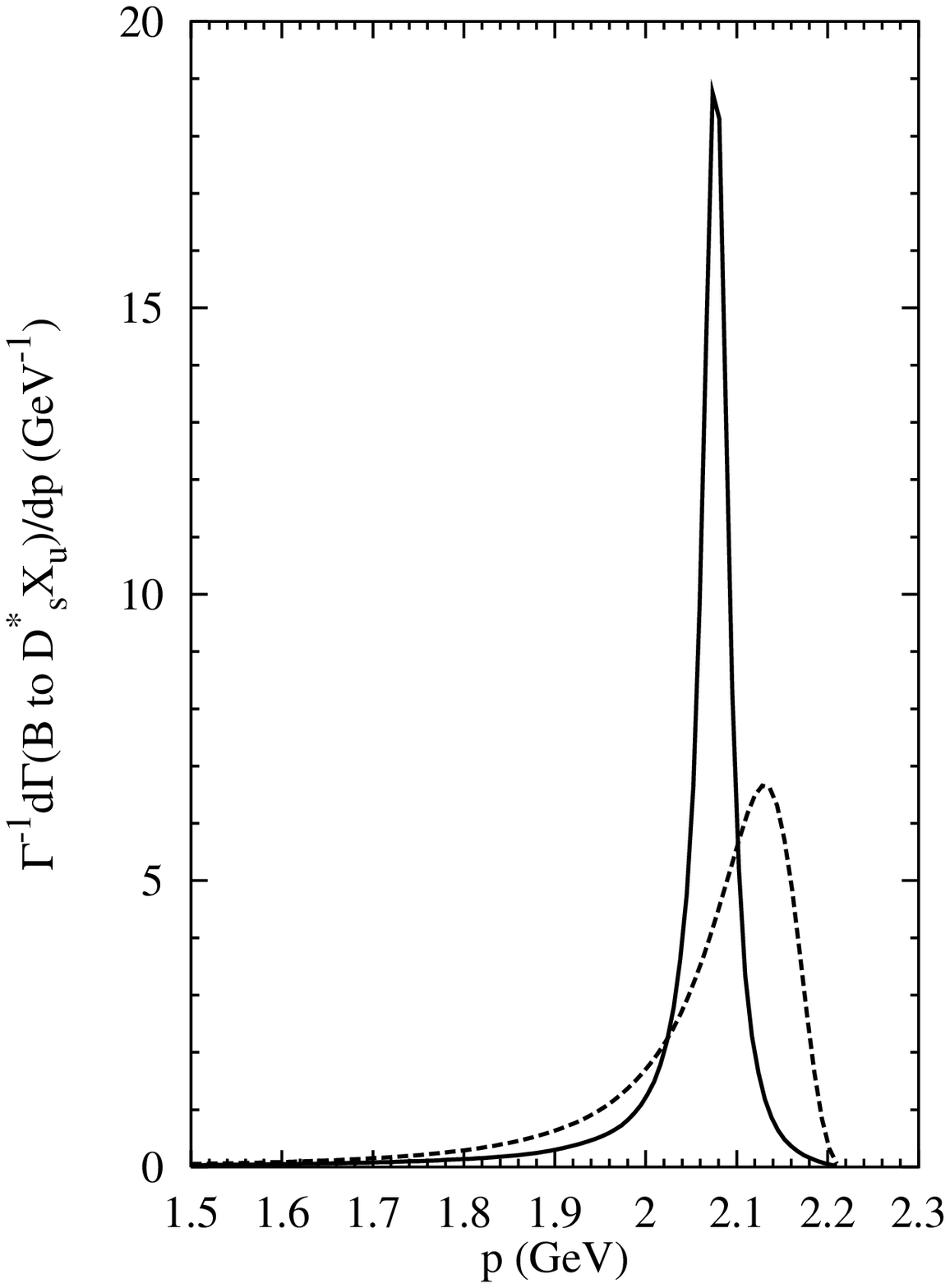}}
{\tighten
\caption {
Momentum spectrum for $D_s^{*-}$ in $\bar B \to D_s^{*-} X_u$.} }
\label{bustar}
\end{figure}

\end{document}